# Larger grains in high-$T_c$ superconductors synthesized by the solid-state reaction route

D.M. Gokhfeld, M.I. Petrov, S.V. Semenov, A.D. Balaev, I.V. Nemtsev, A.D. Vasiliev, M.S. Molokeev

*Kirensky Institute of Physics, Krasnoyarsk Scientific Center,*

*Siberian Branch, Russian Academy of Sciences, Krasnoyarsk, 660036 Russia*

**Abstract** Solid-state synthesis is widely used in exploratory research to study various structural modifications that affect the properties (critical temperature, critical current density, irreversibility field, etc.) of superconductors. The popularity of this method is due to its relative simplicity and availability of the necessary equipment. Combining solid-state synthesis and top-seeded melt growth allows us to increase the grain size in a Tm- and Nd-based 1-2-3 superconductor. Samples with a grain size up to 0.1 mm have been obtained. X-ray diffraction, scanning electron microscopy and magnetization measurements have been used for investigating this superconducting material. The magnetization width $\Delta M$ has increased significantly in the synthesized samples. However the temperature dependence of the intragrain critical current density and the pinning force scaling give evidences that the pinning mechanism in the obtained superconductor is essentially the same as in polycrystalline superconductors synthesized by standard solid-state technology. The increase in grain size in the synthesized samples is the main reason for the high values of $\Delta M$.

## 1. Introduction

An increase in the critical current density of high-$T_c$ superconductors (HTSs) via diverse structural modifications is the subject of many works [1–3]. Polycrystalline superconductors are often employed for such experiments [4–10] since the preparation of HTS crystals is resource intensive and time consuming. Solid-state synthesis is the most popular method for preparing polycrystalline superconductors due to its simplicity. This method allows one to modify the chemical composition and add various admixtures. Material performance, in particular the grain size, is influenced by synthesis temperature, annealing time, pressure, and chemical composition [11–16]. The grain sizes of the 1-2-3 superconductors prepared by solid-state synthesis are few micrometers [13,14,17–20]. The grain size significantly affects the magnetic properties of polycrystalline YBCO [21], e.g. the trapped magnetic field. Smaller particles exhibit specific magnetic behaviors [16], while larger grains potentially enhance superconducting properties because they have higher values of the full penetration field.

A method has previously been proposed for the growth of HTS grains (LSCO [22], YBCO [23,24]) in the mixture of some non-superconducting phases. We modify this method to obtain a large-grain 1-2-3 compound. The main novelty is the use of seed grains as in a top-seeded melt growth technique [25–27].

**2. Experimental**

Two ceramic compounds, $NdBa_2Cu_3O_{7-\delta}$ and $TmBa_2Cu_3O_{7-\delta}$ with peritectic temperatures of 1068 °C and 980 °C respectively, were selected as precursors for the synthesis of a large-grain 1-2-3 compound. These materials were previously synthesized by a conventional solid-state route from $Nd_2O_3$, $Tm_2O_3$, $BaCO_3$, and $CuO$. The synthesized ceramic materials were grinded and mixed. The concentrations of the components were 20 vol.% of $NdBa_2Cu_3O_{7-\delta}$ and 80 vol.% of $TmBa_2Cu_3O_{7-\delta}$, such that the amount of the high-melting phase is below the percolation threshold. The mixture was pressed into tablets and annealed at 980 °C for 1 h. Such annealing leads to the formation of a liquid phase (the product of peritectic decay of $TmBa_2Cu_3O_{7-\delta}$). A low-melting phase can grow around $NdBa_2Cu_3O_{7-\delta}$ grains which have a higher peritectic temperature. The refractory $NdBa_2Cu_3O_{7-\delta}$ phase thus acts as seeds for the growth of the fusible $TmBa_2Cu_3O_{7-\delta}$ phase. After annealing, the sample furnace was cooled at a rate of 0.5° per minute. We label the material as Tm123(Nd123).

The powder diffraction data for Rietveld analysis were collected at room temperature using a Haoyuan DX-2700BH powder diffractometer with Cu-Kα radiation and a linear detector. The step size of 2θ was 0.01°, and the counting time was 0.2 s per step. The scanning electron microscopy (SEM) was performed on a Hitachi TM 4000 Plus microscope. The field dependences of the magnetization $M(H)$ were measured with a Quantum Design MPMS-9 T vibration magnetometer up to 90 kOe. The temperature dependence of the magnetization $M(T)$ was measured with a vibrating magnetometer with a copper coil [28] under zero-field cooled conditions. The sample for magnetic measurements was made in the form of rectangular cuboid with sides of 4.0 mm × 2.6 mm × 3.3 mm (Fig. S1 in Supplementary material). Demagnetization had not been taken into account because it is inconsiderable for polycrystalline superconductors [15].

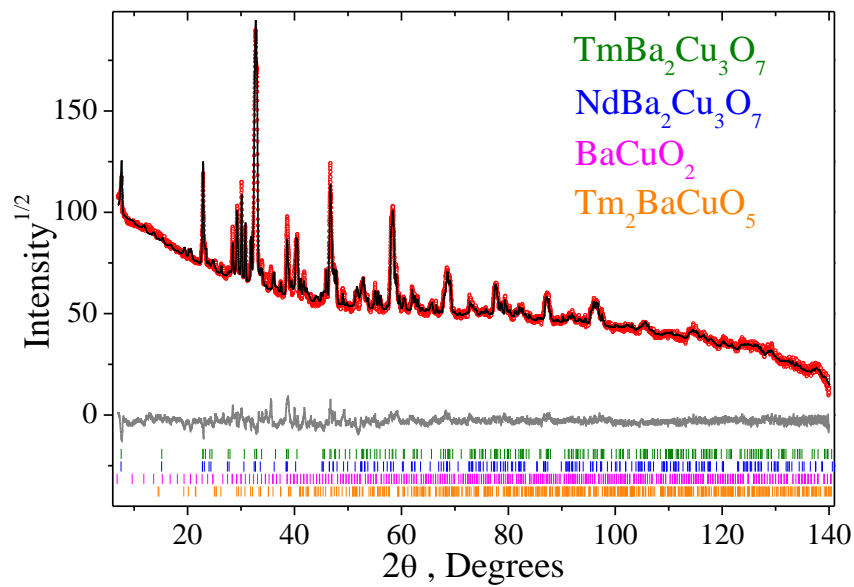

**Fig. 1**. X-ray diffraction data for Tm123(Nd123).

## 3. Results

Figure 1 displays the X-ray diffraction data of Tm123(Nd123). Most of the peaks were indexed by four phases: $TmBa_2Cu_3O_7$, $NdBa_2Cu_3O_7$, $BaCuO_2$, and $Tm_2BaCuO_5$. These structures were hence used as the starting model for Rietveld refinement. The refinements were stable and gave low $R$-factors: $R_{wp}$ = 6.08, $R_p$ = 4.50, and $\chi^2$ = 3.61. Based on the results of the Rietveld refinement, Tm123(Nd123) contains 68 wt.% of the 1-2-3 phase, 18 wt.% of $BaCuO_2$ and 14 wt.% of $Tm_2BaCuO_5$ (green phase). The non-superconducting phases are suspected to be inherited from the precursor ceramics. We assume that the 1-2-3 phase includes $TmBa_2Cu_3O_{7-\delta}$, $NdBa_2Cu_3O_{7-\delta}$, and a continuous series of their solid solutions, $Tm_{(1-x)}Nd_xBa_2Cu_3O_{7-\delta}$.

Figure 2a shows typical micrographs of Tm123(Nd123) obtained using SEM. It is seen that Tm123(Nd123) has many small grains, about 5 μm, and a few grains up to 0.1 mm in size. The size distribution of the grains obtained from a series of micrographs is presented in Figure 2b.

Figure 3 shows the temperature dependence of the magnetization $M(T)$ at an external magnetic field of 10 Oe. The critical temperature $T_c$ = 91.3 K is determined from the beginning of the deviation from the linear high-temperature section on the derivative of $M(T)$. Only one peak on $dM(T)/dT$ (Fig. 3) indicates a unified superconducting phase for this compound.

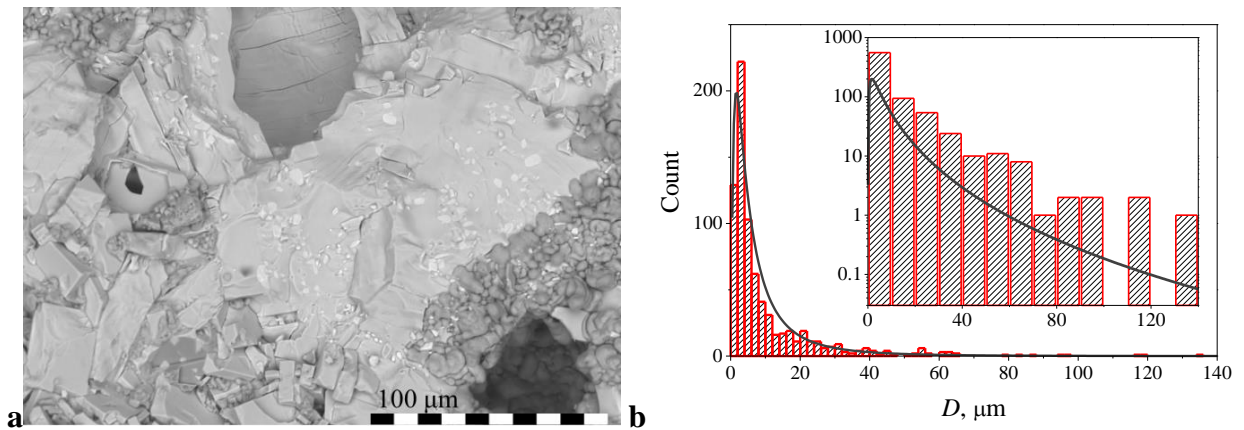

**Fig. 2**. (a) SEM image and (b) the grain size distribution for Tm123(Nd123). Inset shows the same data in a semilogarithmic scale. Solid line is the lognormal distribution.

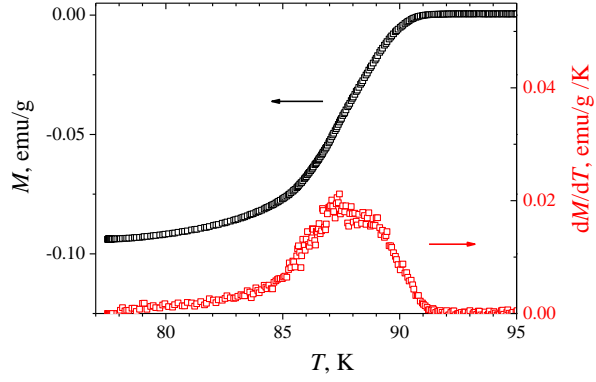

**Fig. 3**. Temperature dependence of the magnetization $M$ and $dM/dT$ for Tm123(Nd123).

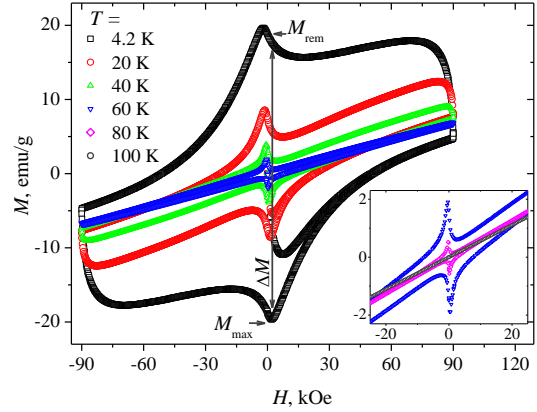

**Fig. 4**. Magnetization hysteresis loops for Tm123(Nd123) at the temperature range of 4.2–100 K. The remanent magnetization $M_{rem}$, the maximum diamagnetic magnetization $M_{max}$, and the magnetization width $\Delta M$ are indicated for the loop at $T = 4.2$ K.

Figure 4 presents the temperature evolution of the magnetic hysteresis loop $M(H)$ for Tm123(Nd123) in the range of 4.2-100 K. The observed temperature evolution of $M(H)$ and the visible paramagnetic contribution are typical for rare-earth element based superconductors (REBCO) [19,29].

## 4. Discussion

The grain sizes in Tm123(Nd123) are described by the lognormal distribution

$$F(D) = \frac{A}{\sigma D/\lambda} \exp\left[-\frac{\ln(D/\lambda)^2}{2\sigma^2}\right].$$

The solid line in Fig. 2b is calculated with the distribution parameters $A = 120$, $\sigma = 1.1$, and $\lambda = 5.5$ μm. The average grain size is 10.4 μm. The largest grains have sizes up to 140 μm. The share of grains with a size greater than 0.1 mm is just 0.4%. However, the volume fraction of such large grains reaches 36%. We assume that these largest grains and have $NdBa_2Cu_3O_{7-d}$ cores and grown $TmBa_2Cu_3O_{7-\delta}$ shells.

The parameter $\Delta M$, the magnetization width, which characterizes the superconducting contribution, for Tm123(Nd123) (Fig. 4) is several times larger than for typical polycrystalline REBCO compounds [13,14,17–19,30]. The precursor superconductors also have values of $\Delta M$ several smaller than this for Tm123(Nd123) [31]. The high value of $\Delta M$ for Tm123(Nd123) may be the result of an increase in the critical current density due to the configuration entropy when the lattice nodes are randomly occupied by Tm and Nd atoms [32].

Let us further consider the critical current density and magnetic flux pinning for Tm123(Nd123). The critical current density $j_c$ for Tm123(Nd123) was estimated from the magnetization hysteresis loops by the formula $j_c = 3\Delta M/D$, where $D$ is the current circulation scale. For polycrystalline samples, these $j_c$ values are the intragrain critical current density since the penetration of Abrikosov vortices into the grains governs the magnetization in large fields and the hysteresis loop [15].

The current circulation scale was determined from the asymmetry of the loop relative to the $M = 0$ axis according to the method described in [33]: $D \approx 2\lambda / [1 - (\Delta M/2|M_{max}|)^{1/3}]$, where $\lambda = 150 \cdot 10^{-9}$ m, $M_{max}$ is the maximal diamagnetic magnetization (Fig. 3b). A value of $D \approx 11$ μm was obtained, which coincides with the average size of the grains in SEM images. The estimated $j_c(H)$ dependencies are shown in Figure 5. The maximum value of $j_c$ is $7.2 \cdot 10^6$ A/cm$^2$ at $T = 4.2$ K and $1.7 \cdot 10^5$ A/cm$^2$ at $T = 80$ K. Since the superconducting phase occupies only 69% of the sample volume, in accordance with the XRD data, the values of $j_c$ for Tm123(Nd123) can be some larger. Accounting the superconducting phase content, $j_c = 1.0 \cdot 10^7$ A/cm$^2$ at $T = 4.2$ K and $j_c = 2.5 \cdot 10^5$ A/cm$^2$ at $T = 80$ K. Comparable values of $j_c$ were observed for polycrystalline YBCO with partial substitutions of Y for rare earth elements (Nd, Ho, Gd) [14,18,19,30]. A single domain REBCO can have smaller values of $j_c$ [34] that is due to chemical inhomogeneity and deficient oxygenation [26,35].

When the external field $H$ is higher than the irreversibility field $H_{irr}(T)$, the magnetization hysteresis loop becomes reversible such that $\Delta M \approx 0$ and $j_c \approx 0$. As it is seen in Fig. 5, the irreversibility field $H_{irr}$ was attained only at $T = 80$ K. So we can plot a curve for the pinning force density $F_p(H) = \mu_0 H \cdot j_c(H)$ at $T = 80$ K (Figure 5b) and apply the Dew-Huge scaling law [36] to determine the pinning mechanism. For high-$T_c$ superconductors the Dew-Huge scaling law can be expressed as $F_p(H) \sim (H/H_{irr})^p \cdot (1-H/H_{irr})^q$, where $p$ and $q$ are positive parameters whose values depend on the type of pinning centers. The data is successfully fitted for $p = 0.8$ and $q = 3.2$ (solid line in Fig. 5b). The observed difference from the Kramer parameters $p = 0.5$, $q = 2$ [37] (dash line in Fig. 5b) is common for high-$T_c$ superconductors [38]. However the maximum of the $F_p(H)$ dependence is located at the same field $H \approx 0.2\ H_{irr}$ as for the Kramer scaling. This may indicate the importance of surface pinning centers [36], namely grain boundaries. The similar scaling parameters were observed for polycrystalline REBCO and YBCO with partial substitutions of Y for rare earth elements [14,18].

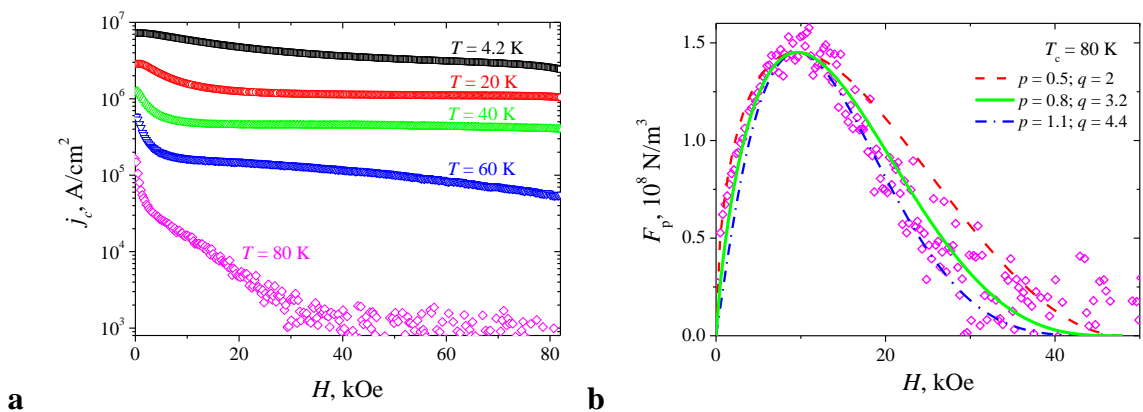

**Fig. 5.** Dependencies of (a) the critical current density $j_c$ and (b) the pinning force density $F_p$ on the magnetic field for Tm123(Nd123). The lines are for the Dew-Huge scaling law (see text).

The change of the maximal critical current density $j_c$ with the temperature is presented in Fig. 6. The data points are obtained from Fig. 5a. The calculated curves corresponding to the main related theoretical models of pinning are also presented in Fig. 6. The curve (1) for pinning from spatial variations in the transition temperature ($\delta T_c$ pinning) is described by $j_c(T) = j_c(0)\ [1-(T/T_c)^2]^{5/2}\ [1+(T/T_c)^2]^{-1/2}$ [39]. The curve (2) for pinning from spatial variations in the charge carrier mean free path ($\delta l$ pinning) is described by $j_c(T) = j_c(0)\ [1-(T/T_c)^2]^{7/6}\ [1+(T/T_c)^2]^{5/6}$ [39]. The curve (3) for strong correlated pinning is described by $j_c(T) = j_c(0)\ \exp(-3(T/T_0)^2)$ [40]. The curve (4) for weak collective pinning is described by $j_c(T) = j_c(0)\ \exp(-T/T_0)$ [41,42]. The $T_0$ parameter corresponds to the characteristic energy of vortex pinning. It is seen that the data points are well fitted by an exponential function (4) associated to the weak collective vortex pinning model. The fitting parameters are $j_c(0) = 8.7 \cdot 10^6$ A/cm$^2$ and $T_0 = 20$ K.

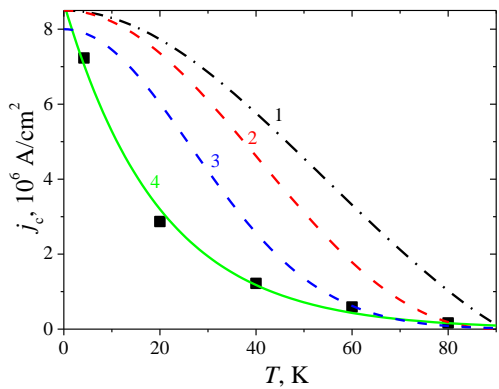

**Fig. 6**. The maximal critical current density for Tm123(Nd123) at different temperatures. Lines are the results of related models (see text).

The analysis of the magnetic flux pinning and the temperature dependence of the critical current density for Tm123(Nd123) shows that the pinning mechanism in Tm123(Nd123) is essentially the same as for polycrystalline superconductors synthesized by standard solid-state technology [14,18,19]. Also it explains similar values of the intragranular critical current density for Tm123(Nd123) and these polycrystalline superconductors. Therefore, the pinning mechanism is not responsible for the high $\Delta M$ value of Tm123(Nd123).

The value of $\Delta M$ can be drastically affected by changing the current circulation scale [15,33]. Figure 7 shows the magnetization hysteresis loop of the Tm123(Nd123) sample at $T = 4.2$ K, the liquid helium temperature. For comparison, the magnetic hysteresis of the polycrystalline superconductor $Y_{0.75}Nd_{0.25}Ba_2Cu_3O_{7-\delta}$ synthesized by standard solid-state technology [18] is also plotted in Figure 7. The hysteresis loops of the samples are noticeably different from each other. The $Y_{0.75}Nd_{0.25}Ba_2Cu_3O_{7-\delta}$ sample has a lower values of the magnetization width $\Delta M$ at any external field $H$ than Tm123(Nd123). The grains in $Y_{0.75}Nd_{0.25}Ba_2Cu_3O_{7-\delta}$ are about 3.8 μm, which is 2.7 times smaller than the average grain size for Tm123(Nd123). It may be claimed that the difference in $\Delta M$ in these samples are mainly due to different grain sizes.

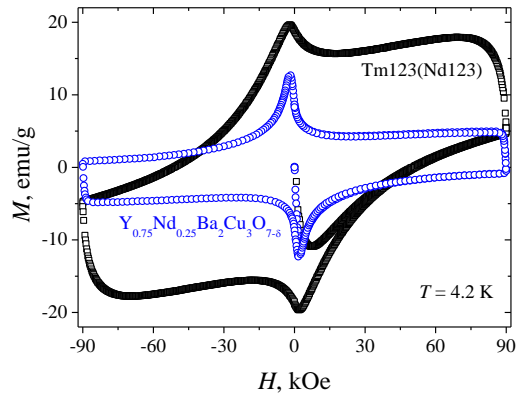

**Fig. 7**. Magnetization hysteresis loops for Tm123(Nd123) and $Y_{0.75}Nd_{0.25}Ba_2Cu_3O_{7-\delta}$ [18].

The remanent magnetization $M_{rem}$, which is a value of $M$ at $H = 0$ as shown in Fig. 3b, is equal to $\Delta M/2$. By definition, $M(H) = -H + B(H)/\mu_0$. Therefore the value of $M_{rem}$ determines the trapped magnetic field $B_{tr} = \mu_0 M_{rem}$ and the trapped magnetic flux $\Phi = B_{tr}S$, where $S$ is the area of the sample cross-section. For the Tm123(Nd123) sample, $B_{tr} = 0.167$ T and $\Phi = 1.35$ μWb are resulted from $M_{rem}$ at $T = 4.2$ K. Polycrystalline 1-2-3 HTSs synthesized using the conventional solid-state routine trap the magnetic field smaller than 0.1 T [18,19,30]. The higher value of $B_{tr}$ in Tm123(Nd123) is also due to the larger size of grains.

In the presented method, the time of annealing seems to be crucial for the growth of the largest grains. The Tm123(Nd123) material annealed for shorter time (15 min) has smaller grains and a smaller values of irreversibility field (see Supplementary material). It is assumed that the average grain size increases with annealing time up to some maximum size. The search for optimal conditions (synthesis time and temperature, precursor concentrations) to increase the volume fraction of large grains and reduce secondary phases is a special technological task for future studies.

## 5. Conclusion

A simple method for producing a REBCO superconductor with a grain size greater than 100 μm has been developed. Annealing of a mixture of Tm123 and Nd123 was used at a temperature in the interval between the peritectic temperatures of the initial components of the mixture. In the resulting material, the large grains compose a long tail of the grain size distribution. The share of grains with a size greater than 0.1 mm is only 0.4% of all grains. At the same time, the volume fraction of these grains reaches 36%. These larger grains are responsible for the record values of $\Delta M$ among polycrystalline high-$T_c$ superconductors. We believe that the proposed method for synthesizing a large-grain superconductor can be further improved by tuning the synthesis time and precursor concentrations.

**Declaration of competing interest**

The authors declare that they have no known competing financial interests or personal relationships that could have appeared to influence the work reported in this paper.

**Acknowledgements**

The XRD, SEM and magnetic measurements were carried out on the equipment of the Krasnoyarsk Regional Center for Collective Use, Krasnoyarsk Scientific Center, Siberian Branch of the Russian Academy of Sciences. This study was carried out within the State assignment of the Ministry of Science and Higher Education of the Russian Federation for the Kirensky Institute of Physics, Siberian Branch of the Russian Academy of Sciences.

**References**

[1]   A. Palau, F. Valles, V. Rouco, M. Coll, Z. Li, C. Pop, B. Mundet, J. Gazquez, R. Guzman, J. Gutierrez, X. Obradors, T. Puig, Disentangling vortex pinning landscape in chemical solution deposited superconducting $YBa_2Cu_3O_{7-x}$ films and nanocomposites, Supercond. Sci. Technol. 31 (2018) 034004. https://doi.org/10.1088/1361-6668/AAA65E.

[2]   S. Eley, A. Glatz, R. Willa, Challenges and transformative opportunities in superconductor vortex physics, J. Appl. Phys. 130 (2021) 050901. https://doi.org/10.1063/5.0055611.

[3] A. Goyal, R. Kumar, H. Yuan, N. Hamada, A. Galluzzi, M. Polichetti, Significantly enhanced critical current density and pinning force in nanostructured, (RE)BCO-based, coated conductor, Nat. Commun. 2024 151. 15 (2024) 1–13. https://doi.org/10.1038/s41467-024-50838-4.

[4] G. Wang, M.J. Raine, D.P. Hampshire, How resistive must grain boundaries in polycrystalline superconductors be, to limit $J_c$?, Supercond. Sci. Technol. 30 (2017) 104001. https://doi.org/10.1088/1361-6668/aa7f24.

[5] A. Öztürk, İ. Düzgün, M. Başoğlu, S. Çelebi, Comparative investigation on magnetic behavior of partial rare earth element (Re: Lu, Yb, and Dy) substituted $Y_{0.5}Re_{0.5}BCO$ (123) superconductors, J. Supercond. Nov. Magn. 33 (2020) 583–590. https://doi.org/10.1007/s10948-019-05233-2.

[6] P. Pęczkowski, P. Zachariasz, C. Jastrzębski, J. Piętosa, E. Drzymała, Ł. Gondek, On the superconductivity suppression in $Eu_{1-x}Pr_xBa_2Cu_3O_{7-\delta}$, Materials. 14 (2021) 3503. https://doi.org/10.3390/ma14133503.

[7] A.L. Pessoa, M.J. Raine, D.P. Hampshire, D.K. Namburi, J.H. Durrell, R. Zadorosny, Successful production of solution blow spun YBCO+Ag complex ceramics, Ceram. Int. 46 (2020) 24097–24101. https://doi.org/10.1016/j.ceramint.2020.06.188.

[8] R. Algarni, E. Hannachi, Y. Slimani, M.A. Almessiere, F. Ben Azzouz, Flux pinning mechanisms of $(YBa_2Cu_3O_{y-d})_{1-x}/(Dy_2O_3)_x$ superconductors (x=0.1 and 0.5 wt%), Ceram. Int. 47 (2021) 6675–6682. https://doi.org/10.1016/j.ceramint.2020.11.007.

[9] S.K. Gadzhimagomedov, D.K. Palchaev, Z.K. Murlieva, M.K. Rabadanov, M.Y. Presnyakov, E. V. Yastremsky, N.S. Shabanov, R.M. Emirov, A.E. Rabadanova, YBCO nanostructured ceramics: Relationship between doping level and temperature coefficient of electrical resistance, J. Phys. Chem. Solids. 168 (2022) 110811. https://doi.org/10.1016/j.jpcs.2022.110811.

[10] A. Raghavan, N.D. Arndt, N. Morales-Colón, E. Wennen, M. Wolfe, C.O. Gandin, K. Nelson, R. Nowak, S. Dillon, K. Sahebkar, R.F. Need, Superconductivity in compositionally complex cuprates with the $YBa_2Cu_3O_{7-x}$ structure, Phys. Rev. Mater. 8 (2024) 024801. https://doi.org/10.1103/PhysRevMaterials.8.024801.

[11] L.E. Richards, H.A. Hoff, P.K. Aggarwal, Grain growth of high $T_c$ $YBa_2Cu_3O_{7-x}$, J. Electron. Mater. 22 (1993) 1233–1239. https://doi.org/10.1007/BF02818066.

[12] Z.D. Yakinci, D.M. Gokhfeld, E. Altin, F. Kurt, S. Altin, S. Demirel, M.A. Aksan, M.E. Yakinci, $J_c$ enhancement and flux pinning of Se substituted YBCO compound, J. Mater. Sci. Mater. Electron. 24 (2013) 4790–4797. https://doi.org/10.1007/s10854-013-1476-8.

[13] P.M. Peczkowski, P. Konieczny, E.M. Dutkiewicz, C. Jastrzebski, P. Zachariasz, E.

Drzymala, A. Zarzycki, D. Bocian, Physico-chemical properties of ceramic high-temperature superconductors with an approximate average radius of rare earth ion(-s) obtained by a solid-state synthesis reaction, SPIE.Proceedings. 11054 (2019) J1–J8. https://doi.org/10.1117/12.2525446.

[14] D.M. Gokhfeld, S. V. Semenov, I. V. Nemtsev, I.S. Yakimov, D.A. Balaev, Magnetic ion substitution and peak effect in YBCO: the strange case of $Y_{1-x}Gd_xBa_2Cu_3O_{7-\delta}$, J. Supercond. Nov. Magn. 35 (2022) 2679–2687. https://doi.org/10.1007/s10948-022-06317-2.

[15] D.M. Gokhfeld, On Estimating the Critical Current Density in Polycrystalline Superconductors Synthesized by Solid-State Method, J. Supercond. Nov. Magn. 36 (2023) 1089–1092. https://doi.org/10.1007/s10948-023-06575-8.

[16] V. Blanco-Gutiérrez, M.J. Torralvo-Fernández, M.A. Alario-Franco, Particle size effect on the superconducting properties of $YBa_2Cu_3O_{7-x}$ particles, Dalt. Trans. 46 (2017) 11698–11703. https://doi.org/10.1039/C7DT01974B.

[17] I.L. Landau, J.B. Willems, J. Hulliger, Detailed magnetization study of superconducting properties of $YBa_2Cu_3O_{7-x}$ ceramic spheres, J. Phys. Condens. Matter. 20 (2008) 095222. https://doi.org/10.1088/0953-8984/20/9/095222.

[18] D.M. Gokhfeld, D.A. Balaev, I.S. Yakimov, M.I. Petrov, S.V. Semenov, Tuning the peak effect in the $Y_{1-x}Nd_xBa_2Cu_3O_{7-\delta}$ compound, Ceram. Int. 43 (2017) 9985–9991. https://doi.org/10.1016/J.CERAMINT.2017.05.011.

[19] D.M. Gokhfeld, S. V. Semenov, K.Y. Terentyev, I.S. Yakimov, D.A. Balaev, Interplay of Magnetic and Superconducting Subsystems in Ho-Doped YBCO, J. Supercond. Nov. Magn. 34 (2021) 2537–2543. https://doi.org/10.1007/s10948-021-05954-3.

[20] U. Erdem, M.B. Turkoz, G. Yıldırım, Y. Zalaoglu, S. Nezir, Refinement of fundamental characteristic properties with homovalent Er/Y partial replacement of $YBa_2Cu_3O_{7-y}$ ceramic matrix, J. Alloys Compd. 884 (2021) 161131. https://doi.org/10.1016/j.jallcom.2021.161131.

[21] A.S. Krasilnikov, L.G. Mamsurova, V.P. Oleshko, N.G. Trusevich, L.G. Shcherbakova, A.A. Vishnev, Effective magnetic grain size and its influence on the intragrain critical current in YBaCuO polycrystals, Supercond. Sci. Technol. 7 (1994) 638. https://doi.org/10.1088/0953-2048/7/9/004.

[22] A.A. Bykov, K.Y. Terent'ev, D.M. Gokhfeld, N.E. Savitskaya, S.I. Popkov, M.I. Petrov, Superconductivity on interfaces of nonsuperconducting granules $La_2CuO_4$ and $La_{1.56}Sr_{0.44}CuO_4$, J. Supercond. Nov. Magn. 31 (2018) 3867–3874. https://doi.org/10.1007/s10948-018-4668-x.

[23] N.A. Kalanda, V.M. Trukhan, S.F. Marenkin, Phase transformations in the systems $Y_2BaCuO_5$-"$Ba_3Cu_5O_8$" and $Y_2BaCuO_5$-$BaCuO_2$, Inorg. Mater. 38 (2002) 597–603. https://doi.org/10.1023/A:1015817603375.

[24] M.I. Petrov, S.I. Popkov, K.Y. Terent'ev, A.D. Vasil'ev, Forming high-temperature superconducting layers at the interfaces between nonsuperconducting phases, Tech. Phys. Lett. 46 (2020) 1004–1007. https://doi.org/10.1134/S1063785020100247.

[25] K. Iida, N.H. Babu, Y. Shi, D.A. Cardwell, Seeded infiltration and growth of large, single domain Y–Ba–Cu–O bulk superconductorswith very high critical current densities, Supercond. Sci. Technol. 18 (2005) 1421. https://doi.org/10.1088/0953-2048/18/11/002.

[26] D.K. Namburi, Y. Shi, D.A. Cardwell, The processing and properties of bulk (RE)BCO high temperature superconductors: current status and future perspectives, Supercond. Sci. Technol. 34 (2021) 053002. https://doi.org/10.1088/1361-6668/ABDE88.

[27] A. Abulaiti, Y. Wan-Min, A novel interior seed addition to improve the levitation force and the trapped field of multi-seeded YBCO bulk superconductors, Supercond. Sci. Technol. 36 (2023) 115010. https://doi.org/10.1088/1361-6668/ACF88C.

[28] A.D. Balaev, Y. V. Boyarshinov, M.M. Karpenko, B.P. Khrustalev, Automated vibration magnetometer with superconducting solenoid, Prib. Tekh. Eksp. 3 (1985) 167.

[29] M. Murakami, N. Sakai, T. Higuchi, S.I. Yoo, Melt-processed light rare earth element-Ba-Cu-O, Supercond. Sci. Technol. 9 (1996) 1015–1032. https://doi.org/10.1088/0953-2048/9/12/001.

[30] H. Theuss, H. Kronmüller, Magnetic properties of $Y_{1-x}Gd_xBa_2Cu_3O_{7-\delta}$ polycrystals, Physica C. 242 (1995) 155–163. https://doi.org/10.1016/0921-4534(94)02404-9.

[31] M.I. Petrov, D.M. Gokhfeld, S.V. Semenov, I.V. Nemtsev, Solid-phase synthesis and properties of a large-grain high-temperature superconductor based on thulium and neodymium, Tech. Phys. Lett. 50 (2024) 11–14. https://doi.org/10.61011/PJTF.2024.17.58573.19956.

[32] A. Yamashita, Y. Shukunami, Y. Mizuguchi, Improvement of critical current density of $REBa_2Cu_3O_{7-\delta}$ by increase in configurational entropy of mixing, R. Soc. Open Sci. 9 (2022) 211874. https://doi.org/10.1098/rsos.211874.

[33] D.M. Gokhfeld, The circulation radius and critical current density in type II superconductors, Tech. Phys. Lett. 45 (2019) 1–3. https://doi.org/10.1134/S1063785019010243.

[34] J. Baumann, Y. Shi, A.R. Dennis, J.H. Durrell, D.A. Cardwell, The influence of porosity on the superconducting properties of Y–Ba–Cu–O single grains, Supercond. Sci. Technol. 36 (2023) 085020. https://doi.org/10.1088/1361-6668/ACE480.

[35] M. Eisterer, S. Haindl, M. Zehetmayer, R. Gonzalez-Arrabal, H.W. Weber, D. Litzkendorf, M. Zeisberger, T. Habisreuther, W. Gawalek, L. Shlyk, G. Krabbes, Limitations for the trapped field in large grain YBCO superconductors, Supercond. Sci. Technol. 19 (2006) 530–536. https://doi.org/10.1088/0953-2048/19/7/S21.

[36] D. Dew-Hughes, Flux pinning mechanisms in type II superconductors, Philos. Mag. 30 (1974) 293–305. https://doi.org/10.1080/14786439808206556.

[37] E.J. Kramer, Scaling laws for flux pinning in hard superconductors, J. Appl. Phys. 44 (1973) 1360–1370. https://doi.org/10.1063/1.1662353.

[38] V. Sandu, Pinning-force scaling and its limitation in intermediate and high temperature superconductors, Mod. Phys. Lett. B. 26 (2012) 1230007. https://doi.org/10.1142/S0217984912300074.

[39] R. Griessen, Wen Hai-Hu, A.J.J. Van Dalen, B. Dam, J. Rector, H.G. Schnack, S. Libbrecht, E. Osquiguil, Y. Bruynseraede, Evidence for mean free path fluctuation induced pinning in $YBa_2Cu_3O_7$ and $YBa_2Cu_4O_8$ films, Phys. Rev. Lett. 72 (1994) 1910–1913. https://doi.org/10.1103/PhysRevLett.72.1910.

[40] D.R. Nelson, V.M. Vinokur, Boson localization and correlated pinning of superconducting vortex arrays, Phys. Rev. B. 48 (1993) 13060. https://doi.org/10.1103/PhysRevB.48.13060.

[41] G. Blatter, M. V. Feigel'man, V.B. Geshkenbein, A.I. Larkin, V.M. Vinokur, Vortices in high-temperature superconductors, Rev. Mod. Phys. 66 (1994) 1125–1388. https://doi.org/10.1103/RevModPhys.66.1125.

[42] T. Puig, J. Gutiérrez, A. Pomar, A. Llordés, J. Gzquez, S. Ricart, F. Sandiumenge, X. Obradors, Vortex pinning in chemical solution nanostructured YBCO films, Supercond. Sci. Technol. 21 (2008) 034008. https://doi.org/10.1088/0953-2048/21/3/034008.